\documentclass[hyper]{JHEP} 
\usepackage{graphicx,amssymb,bm,latexsym,amsmath}
\newcommand\fverb{\setbox\pippobox=\hbox\bgroup\verb}
\newcommand\fverbdo{\egroup\medskip\noindent%
            \fbox{\unhbox\pippobox}\ }

\newcommand\fverbit{\egroup\item[\fbox{\unhbox\pippobox}]}
\newbox\pippobox
\title{Rotating and Orbiting Strings in Dp-brane background}
\author{Sagar Biswas and Kamal L. Panigrahi\\
Department of Physics, \\
Indian Institute of Technology Kharagpur,\\
Kharagpur-721 302, INDIA \\
Email: \email{sbiswas,panigrahi@phy.iitkgp.ernet.in}} \vskip .2in
\abstract{We probe the open fundamental strings in Dp-brane (p=1,
3, 5) backgrounds and find new classes of rotating and orbiting
string solutions. We show that for various worldsheet embedding ansatz we get
solutions of the string equations of motion that correspond to
the well known giant magnon and single spikes, in addition to few
new solutions corresponding to the orbiting strings. We make a
systematic study of both rigidly rotating and orbiting strings in
D1, D3 and D5-brane backgrounds.} \keywords{AdS-CFT correspondence, Bosonic
Strings}

\begin{document}
\section{Introduction and Summary}
The celebrated AdS/CFT duality relates weakly coupled type IIB
string theory on the $AdS_5 \times S^5$ to strongly coupled large
$N$ $ \mathcal{N} = 4$ super Yang-Mills (SYM) theory in 3+1
dimensional spacetime {\cite{Maldacena:1997re}},
{\cite{Gubser:1998bc}}, {\cite{Witten:1998qj}}. Proving the
conjecture beyond supergravity approximations is quite a hard
problem because of the unavailability of fully quantized string
spectrum in the AdS side. However over the past few years there
has been lot of attempts in understanding the duality in various
sub sectors of the theories. It has been noticed that in large
angular momentum region one can use the semiclassical
approximation to find the string spectrum and eventually look for
the string state-operator matching more precisely {\cite{Berenstein:2002jq}}, {\cite{Gubser:2002tv}}. With the
further realization that the counting of gauge invariant operators
from gauge theory side can be elegantly formulated in terms of an
integrable spin chain {\cite{Minahan:2002ve}}, {\cite{Kazakov:2004qf}}, {\cite{Beisert:2004hm}}, {\cite{Arutyunov:2004vx}}, {\cite{Beisert:2010jr}}, it has been established that integrability
played an important role on both sides of the duality, since the
dual string theory is integrable in the semiclassical limit. In
this context, in {\cite{Hofman:2006xt}}, a special limit was put
forth using which both sides of the duality was analyzed in
detail. In particular, the spectrum on the field theory side was
shown to be consists of an elementary excitation, the so called
magnon which carries a momentum $p$ along the finitely or
infinitely long spin chain. The dual string state, derived from
the rigidly rotating string in the ${\mathbb {R}} \times S^3$
presenting the same dispersion relation in the large 't Hooft
limit, is known as the giant magnon. A more general class of
rigidly rotating string solutions were also proposed in
\cite{Kruczenski:2004wg} which are dual to a higher twist
operators in the boundary field theory. These kind of solutions
are called spiky strings. In addition to the rigidly rotating
strings, the spinning and pulsating strings have also been found
out to have exact correspondence with some dual operators in the
gauge theory {\cite{Engquist:2003rn}}, {\cite{Smedback:1998yn}}. Since
then a fairly large class of rotating and pulsating string
solutions in various asymptotically AdS and non-AdS backgrounds
have been studied and the dual operators have also been examined
carefully over the past few years {\cite{Armoni:2002xp}} -  {\cite{Banerjee:2014rza}} .

In the recent past, the integrability of the classical string
motion in curved D$p$-brane background has been explored in
{\cite{Stepanchuk:2012xi}} in an attempt to understand the integrability of
the full string equations of motion. It was shown that though the
point like string equations are integrable, the equations
describing an extended string in the complete D-brane background
are not. We wish to further shed light on this, by probing the
fundamental string in various Dp-branes of type IIB string theory
and look for various rigidly rotating and orbiting string
solutions. It is is well known that among various Dp branes only
D3 brane is integrable in the near horizon limit as it produces
$AdS_5 \times S^5$ geometry. We have found the known giant magnons
and spikes like solutions for the string in semiclassical limit,
whose dual operators are well identified. Along with these, we
find out other kind of rotating string solutions in the D3-brane
background. Infact both kind of solutions are obtained by looking
at consistent worldsheet embedding. Moreover, we find out new orbiting
solutions for the string in D3-brane background.

We also try to find some rigidly rotating and orbiting string
solutions in D5 and D1 brane background. The string solutions in
these backgrounds are non integrable in its full generalization as
studied in {\cite{Stepanchuk:2012xi}}, but we manage to extract some nice
results by considering different subspaces of the full space. Such
an attempt was done in {\cite{Banerjee:2014rza}} in the context of
the so called supergravity puff field theory background. The
reason why one should expect to get some viable solution by
considering a particular subspace can be understood as follows. We
choose our ansatz in such a way that all non isometry coordinates
becomes variables. For general background, by solving the string
equation of motions, one get highly coupled and nonlinear
differential equations for these variables which are hard to
solve. But, by looking at the Virasoro one can get an idea that
some kind of solution is possible if we keep only one variable and
put the rest as constants. We can choose different types of
subspaces. As the variables corresponds to non-isometric
coordinates, putting them as constants will impose severe non
trivial constraints on the system, which can be easily obtained
from the corresponding equation of motion. In other words putting
these variables as constant are the solutions of the equation of
motion provided that they satisfy the constraint equations. So,
the solutions we present for a particular subspace are valid in
some particular constraint space. The rest of the paper has been
organised as follows. In section 2 we write the most general
string equation of motion and Virasoro constraints in near horizon
Dp-brane backrgound. Section 3 is devoted to the study of F-string
in the D3-brane background. In addition to finding the usual giant
magnon and single spike solutions we find few general rotating
strings with several angular momenta and further the orbiting
solutions. We comments of the possible field theory operators they
might correspond to. In section-4 we study the solutions to the
string equations of motion in the D5-brane background and find a
large class of solutions. In section-5 we study the rigidly
rotating and orbiting string solutions in D1-brane background.
Finally in section 6 we present our conclusions with some remarks.

\section{Probing the fundamental string in Dp-brane backgrounds}
The low energy background field solutions describing a stack of
$k$ Dp-branes of ten-dimensional supergravity is given by
\begin{eqnarray}
ds^2 &=& h^{-\frac{1}{2}}(r)\Big[-dt^2 + \sum_{i=1}^{p}dx_i^2\Big]
+ h^{\frac{1}{2}}(r) \sum_{i=p+1}^{9}dy_i^2, \nonumber \\ e^{\phi}
&=& h^{\frac{3-p}{4}}, \>\>\> C_{0\cdots p} = 1-\frac{1}{h(r)} \ ,
\label{metric 1}
\end{eqnarray}
where $C_{p+1}$ is the ($p+1$)- dimensional Ramond-Ramond (RR)
form which couples to Dp-brane, and $\phi$ is the dilaton. When $p
< 7$ the harmonic function $h(r)$ is explicitly given by
$h(r)=1+\frac{k}{r^{7-p}}$ with $r^2 = y^2_{p+1} + \cdots +
dy^2_9$ and $k=c_pg_sNl_s^{7-p}$ being  $c_p=(2\sqrt{\pi})^{5-p}
\Gamma(\frac{7-p}{2})$. The transverse space part of the metric
(\ref{metric 1}) can be written as,
\begin{eqnarray}
    \sum_{i=p+1}^{9} dy_i^2 &=& dr^2 + r^2d\Omega^2_{8-p} = dr^2
+ r^2(d\phi_1^2 + \sin^2\phi_1 d\phi_2^2 + \cos^2\phi_1d\Omega^2_{6-p})
\nonumber \\
&=& dr^2 + r^2[d\phi_1^2 + \sin^2\phi_1 d\phi_2^2 + \cos^2\phi_1(d\phi_3^2
+ \sin^2\phi_3d\phi_4^2 + \cos^2\phi_3d\Omega^2_{4-p})] \nonumber \\
\end{eqnarray}
We will be interested in Dp-brane $(p = 1, 3, 5)$ backgrounds, so
one should stop when the suffix of $\Omega$ becomes $2-p$. Now we
can rewrite the metric for Dp-branes, $(p =1, 3, 5)$,
\begin{eqnarray}
ds^2 &=& h^{-\frac{1}{2}}(r)\Big[-dt^2 + \sum_{i=1}^{p}dx_i^2\Big]
+ h^{\frac{1}{2}}(r) \Big[dr^2 + r^2\{d\phi_{2-p}^2 +
\sin^2\phi_{2-p} d\phi^2_{3-p} \nonumber \\ &+&
\cos^2\phi_{2-p}(d\phi_{4-p}^2 + \sin^2\phi_{4-p}d\phi_{5-p}^2 +
\cos^2\phi_{4-p}(d\phi_{6-p}^2 \nonumber \\ &+&
\sin^2\phi_{6-p}d\phi_{7-p}^2 + \cos^2\phi_{6-p}d\phi_{8-p}^2 ))\}
\Big], \nonumber \\ e^{\phi} &=& h^{\frac{3-p}{4}}, \>\>\>
C_{0\cdots p} = 1-\frac{1}{h(r)} \ . \label{metric 2}
\end{eqnarray}
In near horizon limit $r\rightarrow 0$, the
metric, dilaton and RR $p$-form field become,
\begin{eqnarray}
ds^2 &=& \frac{r^{\frac{7-p}{2}}}{\sqrt{k}}\Big[-dt^2 +
\sum_{i=1}^{p}dx_i^2\Big] + \frac{\sqrt{k}}{r^{\frac{7-p}{2}}}
\Big[dr^2 + r^2\{d\phi_{2-p}^2 + \sin^2\phi_{2-p} d\phi^2_{3-p}
\nonumber \\ &+& \cos^2\phi_{2-p}(d\phi_{4-p}^2 +
\sin^2\phi_{4-p}d\phi_{5-p}^2 + \cos^2\phi_{4-p}(d\phi_{6-p}^2
\nonumber \\ &+& \sin^2\phi_{6-p}d\phi_{7-p}^2 +
\cos^2\phi_{6-p}d\phi_{8-p}^2 ))\} \Big], \nonumber \\ e^{\phi}
&=& \Big(\frac{k}{r^{7-p}}\Big)^{\frac{3-p}{4}}, \>\>\> C_{0\cdots
p} = 1-\frac{r^{7-p}}{k} \ . \label{metric 3}
\end{eqnarray}
Now rescaling $t \rightarrow \sqrt{k}t$ and $x_i \rightarrow
\sqrt{k} x_i$, the metric in equation (\ref{metric 3}) becomes,
\begin{eqnarray}
ds^2 &=& \sqrt{k} r^{\frac{7-p}{2}}\Big[-dt^2 +
\sum_{i=1}^{p}dx_i^2\Big] + \frac{\sqrt{k}}{r^{\frac{7-p}{2}}}
\Big[dr^2 +
r^2\{d\phi_{2-p}^2 + \sin^2\phi_{2-p} d\phi^2_{3-p} \nonumber \\
&+& \cos^2\phi_{2-p}(d\phi_{4-p}^2 + \sin^2\phi_{4-p}d\phi_{5-p}^2
+ \cos^2\phi_{4-p}(d\phi_{6-p}^2 \nonumber \\ &+&
\sin^2\phi_{6-p}d\phi_{7-p}^2 + \cos^2\phi_{6-p}d\phi_{8-p}^2 ))\}
\Big] \ , \label{metric 4}
\end{eqnarray}
The Polyakov action of the F-string in the background (\ref{metric
4}) is given by,
\begin{equation}
S=-\frac{1}{4\pi\alpha^{\prime}}\int d\sigma d\tau
[\sqrt{-\gamma}\gamma^{\alpha \beta}g_{MN}\partial_{\alpha} X^M
\partial_{\beta}X^N ]\ ,
\end{equation}
where $\gamma^{\alpha \beta}$ is the world-sheet metric . In
conformal gauge (i.e. $\sqrt{-\gamma}\gamma^{\alpha
\beta}=\eta^{\alpha \beta}$) with $\eta^{\tau \tau}=-1$,
$\eta^{\sigma \sigma}=1$ and $\eta^{\tau \sigma}=\eta^{\sigma
\tau}=0$, the Polyakov action in the above background takes the
form,
\begin{eqnarray}
S &=& -\frac{\sqrt{k}}{4\pi}\int d\sigma
d\tau\Big[r^{\frac{7-p}{2}}\{-({t^{\prime}}^2-\dot{t}^2) +
{x_i^{\prime}}^2-\dot{x_i}^2\} + \frac{1}{r^{\frac{7-p}{2}}}({r^{\prime}}^2-\dot{r}^2)
 \nonumber \\
&+& r^{\frac{p-3}{2}}\{({\phi_{2-p}^{\prime}}^2 -
\dot{\phi}_{2-p}^2) +  \sin^2\phi_{2-p}({\phi_{3-p}^{\prime}}^2 -
\dot{\phi}_{3-p}^2) +  \cos^2\phi_{2-p}({\phi_{4-p}^{\prime}}^2 -
\dot{\phi}_{4-p}^2) \nonumber \\ &+&
\cos^2\phi_{2-p}\sin^2\phi_{4-p}({\phi_{5-p}^{\prime}}^2 -
\dot{\phi}_{5-p}^2) +
\cos^2\phi_{2-p}\cos^2\phi_{4-p}({\phi_{6-p}^{\prime}}^2 -
\dot{\phi}_{6-p}^2) \nonumber \\ &+&
\cos^2\phi_{2-p}\cos^2\phi_{4-p}
\sin^2\phi_{6-p}({\phi_{7-p}^{\prime}}^2 - \dot{\phi}_{7-p}^2)
\nonumber \\ &+&
\cos^2\phi_{2-p}\cos^2\phi_{4-p}\cos^2\phi_{6-p}({\phi_{8-p}^{\prime}}^2
- \dot{\phi}_{8-p}^2)\}\Big] \ ,
\end{eqnarray}
where `dots' and `primes' denote the derivative with respect to
$\tau$ and $\sigma$ respectively. Now we choose our ansatz in such
a way that the non isometry coordinates become variables,
\begin{eqnarray}
t &=& \tau + h_0(y) ,\>\> x_i=\nu_i(\tau+h_i(y)),\>\>\>
r=r(y),\nonumber \\ \phi_{2-p} &=& \phi_{2-p}(y), \>\>\>
\phi_{3-p} = \omega_{3-p}(\tau+g_{3-p}(y)), \>\>\>
\phi_{4-p}=\phi_{4-p}(y), \nonumber \\  \phi_{5-p} &=&
\omega_{5-p}(\tau + g_{5-p}(y)), \>\>\> \phi_{6-p}=\phi_{6-p}(y),
\nonumber \\ \phi_{7-p} &=& \omega_{7-p}(\tau + g_{7-p}(y)),\>\>\>
\phi_{8-p} = \omega_{8-p}(\tau + g_{8-p}(y)) \ , \label{ansatz 1}
\end{eqnarray}
where $y=\sigma-v\tau$ and $i=1,\cdots, p$. Variation of the
action with respect to $X^M$ gives us the following equations of
motion
\begin{eqnarray}
2\partial_{\alpha}(\eta^{\alpha \beta} \partial_{\beta}X^Ng_{KN})
&-& \eta^{\alpha \beta} \partial_{\alpha} X^M \partial_{\beta}
X^N\partial_K g_{MN} =0 \ ,
\end{eqnarray}
and variation with respect to the metric gives the two Virasoro
constraints,
\begin{eqnarray}
g_{MN}(\partial_{\tau}X^M \partial_{\tau}X^N +
\partial_{\sigma}X^M \partial_{\sigma}X^N)&=&0 \ , \nonumber \\
g_{MN}(\partial_{\tau}X^M \partial_{\sigma}X^N)&=&0 \ .
\end{eqnarray}
Next we have to solve these equations by the ansatz we have
proposed above in (\ref{ansatz 1}). Solving for the isometry
coordinates $t$, $x_i$, $\phi_{3-p}$, $\phi_{5-p}$,
$\phi_{7-p}$ and $\phi_{8-p}$ respectively, we get,
\begin{eqnarray}
\frac{\partial h_0}{\partial y} &=&
\frac{1}{1-v^2}[r^{\frac{p-7}{2}}c_0 - v], \nonumber \\
\frac{\partial h_i}{\partial y} &=&
\frac{1}{1-v^2}[r^{\frac{p-7}{2}}c_i - v], \nonumber \\
\frac{\partial g_{3-p}}{\partial y} &=&
\frac{1}{1-v^2}\Big[\frac{d_{3-p}}{r^{\frac{p-3}{2}}
\sin^2\phi_{2-p}} - v\Big], \nonumber \\  \frac{\partial
g_{5-p}}{\partial y} &=&
\frac{1}{1-v^2}\Big[\frac{d_{5-p}}{r^{\frac{p-3}{2}}
\cos^2\phi_{2-p} \sin^2\phi_{4-p}} - v\Big], \nonumber \\
\frac{\partial g_{7-p}}{\partial y} &=&
\frac{1}{1-v^2}\Big[\frac{d_{7-p}}{r^{\frac{p-3}{2}}
\cos^2\phi_{2-p} \cos^2\phi_{4-p}\sin^2\phi_{6-p}} - v\Big],
\nonumber \\ \frac{\partial g_{8-p}}{\partial y} &=&
\frac{1}{1-v^2}\Big[\frac{d_{8-p}}{r^{\frac{p-3}{2}}
\cos^2\phi_{2-p} \cos^2\phi_{4-p}\cos^2\phi_{6-p}} - v\Big] \ ,
\label{eom 1}
\end{eqnarray}
where $c_0$, $c_i$  ($i=1,\cdots,p$), $d_{3-p}$, $d_{5-p}$,
$d_{7-p}$ and $d_{8-p}$ are integration constants. Now we want to
solve the equations of motion corresponding to the non isometry
coordinates one by one. First for $\phi_{6-p}$ we get,
\begin{eqnarray}
&& (1-v^2)^2\frac{\partial}{\partial y}\Big[r^{\frac{p-3}{2}}
\cos^2\phi_{2-p}\cos^2\phi_{4-p}\frac{\partial
\phi_{6-p}}{\partial y}\Big] \nonumber \\ &=&
r^{\frac{p-3}{2}}\cos^2\phi_{2-p}\cos^2\phi_{4-p}
\sin\phi_{6-p}\cos\phi_{6-p}
\Big[\frac{\omega_{7-p}^2d_{7-p}^2}{r^{p-3}\cos^4\phi_{2-p}
\cos^4\phi_{4-p}\sin^4\phi_{6-p}} \nonumber \\ &-&
\frac{\omega_{8-p}^2d_{8-p}^2}{r^{p-3}\cos^4\phi_{2-p}\cos^4\phi_{4-p}
\cos^4\phi_{6-p}} + \omega_{8-p}^2 - \omega_{7-p}^2 \Big] \ .
\label{eom 2}
\end{eqnarray}
For $\phi_{4-p}$ we get,
\begin{eqnarray}
&& (1-v^2)^2\frac{\partial}{\partial y}\Big[r^{\frac{p-3}{2}}
\cos^2\phi_{2-p}\frac{\partial \phi_{4-p}}{\partial y}\Big] =
r^{\frac{p-3}{2}}\cos^2\phi_{2-p}\sin\phi_{4-p}\cos\phi_{4-p}
\nonumber \\ &\times&
\Big[\frac{\omega_{5-p}^2d_{5-p}^2}{r^{p-3}\cos^4\phi_{2-p}
\sin^4_{4-p}} -
\frac{\omega_{7-p}^2d_{7-p}^2}{r^{p-3}\cos^4\phi_{2-p}
\cos^4\phi_{4-p}\sin^2\phi_{6-p}} \nonumber \\ &-&
\frac{\omega_{8-p}^2d_{8-p}^2}{r^{p-3}\cos^4\phi_{2-p}\cos^4\phi_{4-p}
\cos^2\phi_{6-p}} - \omega_{5-p}^2 +
\omega_{7-p}^2\sin^2\phi_{6-p} \nonumber \\ &+&
\omega_{8-p}^2\cos^2\phi_{6-p}   - (1-v^2)^2 \Big(\frac{\partial
\phi_{6-p}}{\partial y}\Big)^2  \Big] \ . \label{eom 3}
\end{eqnarray}
For $\phi_{2-p}$ on the other hand we get,
\begin{eqnarray}
&& (1-v^2)^2\frac{\partial}{\partial y}\Big[r^{\frac{p-3}{2}}
\frac{\partial \phi_{2-p}}{\partial y}\Big] =
r^{\frac{p-3}{2}}\sin\phi_{2-p}\cos\phi_{2-p} \Big[
\frac{\omega_{3-p}^2d_{3-p}^2}{r^{p-3}\sin^4\phi_{2-p}} \nonumber
\\ &-&
\frac{\omega_{5-p}^2d_{5-p}^2}{r^{p-3}\cos^4\phi_{2-p}\sin^2_{4-p}}
-
\frac{\omega_{7-p}^2d_{7-p}^2}{r^{p-3}\cos^4\phi_{2-p}\cos^2\phi_{4-p}
\sin^2\phi_{6-p}} \nonumber \\ &-&
\frac{\omega_{8-p}^2d_{8-p}^2}{r^{p-3}
\cos^4\phi_{2-p}\cos^2\phi_{4-p}\cos^2\phi_{6-p}} - \omega_{3-p}^2
+ \omega_{5-p}^2\sin^2\phi_{4-p} \nonumber \\ &+&
\omega_{7-p}^2\cos^2\phi_{4-p}\sin^2\phi_{6-p} +
\omega_{8-p}^2\cos^2\phi_{4-p}\cos^2\phi_{6-p} \nonumber \\ &-&
(1-v^2)^2\Big( \Big(\frac{\partial \phi_{4-p}}{\partial y}\Big)^2
+ \cos^2\phi_{4-p}\Big(\frac{\partial \phi_{6-p}}{\partial
y}\Big)^2  \Big) \Big] \ . \label{eom 4}
\end{eqnarray}
Finally solving for $r$ we get,
\begin{eqnarray}
&& 2(1-v^2)^2\frac{\partial}{\partial y} \Big[r^{\frac{p-7}{2}}
\frac{\partial r}{\partial y}\Big] = [1-c_0^2r^{p-7} - \nu_i^2\{1
- c_i^2r^{p-7}\}]\partial_r(r^{\frac{7-p}{2}}) \nonumber \\ &+&
(1-v^2)^2\Big(\frac{\partial r}{\partial
y}\Big)^2\partial_r(r^{\frac{p-7}{2}}) +
\Big[\frac{\omega_{3-p}^2d_{3-p}^2}{r^{p-3}\sin^2\phi_{2-p}} +
\frac{\omega_{5-p}^2d_{5-p}^2}{r^{p-3}\cos^2\phi_{2-p}\sin^2_{4-p}}
\nonumber \\ &+&
\frac{\omega_{7-p}^2d_{7-p}^2}{r^{p-3}\cos^2\phi_{2-p}\cos^2\phi_{4-p}
\sin^2\phi_{6-p}} + \frac{\omega_{8-p}^2d_{8-p}^2}{r^{p-3}
\cos^2\phi_{2-p}\cos^2\phi_{4-p}\cos^2\phi_{6-p}} \nonumber \\ &-&
\omega_{3-p}^2\sin^2\phi_{2-p} -
\omega_{5-p}^2\cos^2\phi_{2-p}\sin^2\phi_{4-p} -
\omega_{7-p}^2\cos^2\phi_{2-p}\cos^2\phi_{4-p}\sin^2\phi_{6-p}
\nonumber \\ &-&
\omega_{8-p}^2\cos^2\phi_{2-p}\cos^2\phi_{4-p}\cos^2\phi_{6-p} +
(1-v^2)^2\Big( \Big(\frac{\partial \phi_{2-p}}{\partial y}\Big)^2
+ \cos^2\phi_{2-p}\Big(\frac{\partial \phi_{4-p}}{\partial
y}\Big)^2 \nonumber \\ &+&
\cos^2\phi_{2-p}\cos^2\phi_{4-p}\Big(\frac{\partial
\phi_{6-p}}{\partial y}\Big)^2
\Big)\Big]\partial_r(r^{\frac{p-3}{2}})\ . \nonumber \\ \label{eom
5}
\end{eqnarray}
Clearly these are a set of highly coupled nonlinear equations. It
will be hard to solve this system of equations in full generality.
Now let's see what we get from Virasoro constraints.

The Virasoro constraint $g_{MN}(\partial_{\tau}X^M
\partial_{\sigma}X^N)=0$ is given by,
\begin{eqnarray}
&& (1 - v^2)^2 \Big[\Big(\frac{\partial r}{\partial y}\Big)^2 +
r^2\Big(\frac{\partial \phi_{2-p}}{\partial y}\Big)^2 +
r^2\cos^2\phi_{2-p}\Big(\frac{\partial \phi_{4-p}}{\partial
y}\Big)^2  \nonumber \\ &+&
r^2\cos^2\phi_{2-p}\cos^2\phi_{4-p}\Big(\frac{\partial
\phi_{6-p}}{\partial y}\Big)^2 \Big] = (1-\nu_i^2)r^{7-p} + c_0^2
- \nu_i^2c_i^2  - r^2\omega_{3-p}^2\sin^2\phi_{2-p} \nonumber \\
&-& r^2\omega_{5-p}^2\cos^2\phi_{2-p}\sin^2\phi_{4-p} -
r^2\omega_{7-p}^2\cos^2\phi_{2-p}\cos^2\phi_{4-p}\sin^2\phi_{6-p}
\nonumber \\ &-&
r^2\omega_{8-p}^2\cos^2\phi_{2-p}\cos^2\phi_{4-p}\cos^2\phi_{6-p}
- \frac{\omega_{3-p}^2d_{3-p}^2}{r^{p-5}\sin^2\phi_{2-p}} -
\frac{\omega_{5-p}^2d_{5-p}^2}{r^{p-5}\cos^2\phi_{2-p}
\sin^2\phi_{4-p}} \nonumber \\ &-&
\frac{\omega_{7-p}^2d_{7-p}^2}{r^{p-5}\cos^2\phi_{2-p}\cos^2\phi_{4-p}
\sin^2\phi_{6-p}} - \frac{\omega_{8-p}^2d_{8-p}^2}{r^{p-5}
\cos^2\phi_{2-p}\cos^2\phi_{4-p} \cos^2\phi_{6-p}}  \nonumber \\
&+& \frac{1+v^2}{v}r^{\frac{7-p}{2}}(-c_0 + \nu_i^2c_i +
\omega_{3-p}^2d_{3-p} + \omega_{5-p}^2d_{5-p} +
\omega_{7-p}^2d_{7-p} + \omega_{8-p}^2d_{8-p}) \ .
\label{Virasoro 1}
\end{eqnarray}
Again the Virasoro $g_{MN}(\partial_{\tau}X^M \partial_{\tau}X^N +
\partial_{\sigma}X^M \partial_{\sigma}X^N)=0$ becomes,
\begin{eqnarray}
&& (1 - v^2)^2 \Big[\Big(\frac{\partial r}{\partial y}\Big)^2 +
r^2\Big(\frac{\partial \phi_{2-p}}{\partial y}\Big)^2 +
r^2\cos^2\phi_{2-p}\Big(\frac{\partial \phi_{4-p}}{\partial
y}\Big)^2  \nonumber \\ &+&
r^2\cos^2\phi_{2-p}\cos^2\phi_{4-p}\Big(\frac{\partial
\phi_{6-p}}{\partial y}\Big)^2 \Big] = (1-\nu_i^2)r^{7-p} + c_0^2
- \nu_i^2c_i^2  - r^2\omega_{3-p}^2\sin^2\phi_{2-p} \nonumber \\
&-& r^2\omega_{5-p}^2\cos^2\phi_{2-p}\sin^2\phi_{4-p} -
r^2\omega_{7-p}^2\cos^2\phi_{2-p}\cos^2\phi_{4-p}\sin^2\phi_{6-p}
\nonumber \\ &-&
r^2\omega_{8-p}^2\cos^2\phi_{2-p}\cos^2\phi_{4-p}\cos^2\phi_{6-p}
- \frac{\omega_{3-p}^2d_{3-p}^2}{r^{p-5}\sin^2\phi_{2-p}} -
\frac{\omega_{5-p}^2d_{5-p}^2}{r^{p-5}\cos^2\phi_{2-p}
\sin^2\phi_{4-p}} \nonumber \\ &-&
\frac{\omega_{7-p}^2d_{7-p}^2}{r^{p-5}\cos^2\phi_{2-p}\cos^2\phi_{4-p}
\sin^2\phi_{6-p}} - \frac{\omega_{8-p}^2d_{8-p}^2}{r^{p-5}
\cos^2\phi_{2-p}\cos^2\phi_{4-p} \cos^2\phi_{6-p}}  \nonumber \\
&+& \frac{4v}{(1+v^2)}r^{\frac{7-p}{2}}(-c_0 + \nu_i^2c_i +
\omega_{3-p}^2d_{3-p} + \omega_{5-p}^2d_{5-p} +
\omega_{7-p}^2d_{7-p} + \omega_{8-p}^2d_{8-p}) \ .
\label{Virasoro 2}
\end{eqnarray}
By subtracting the two Virasoro constraints we get the following
relation,
\begin{equation}
-c_0 + \nu_i^2c_i + \omega_{3-p}^2d_{3-p} + \omega_{5-p}^2d_{5-p}
+ \omega_{7-p}^2d_{7-p} + \omega_{8-p}^2d_{8-p} =0 \ .
\label{constraint 1}
\end{equation}
If we demand that the constraint (\ref{constraint 1}) is always
satisfied by the solution we are going to present, then we can put
this constraint back into the Virasoro, and we obtain,
\begin{eqnarray}
&& (1 - v^2)^2 \Big[\Big(\frac{\partial r}{\partial y}\Big)^2 +
r^2\Big(\frac{\partial \phi_{2-p}}{\partial y}\Big)^2 +
r^2\cos^2\phi_{2-p}\Big(\frac{\partial \phi_{4-p}}{\partial
y}\Big)^2  \nonumber \\ &+&
r^2\cos^2\phi_{2-p}\cos^2\phi_{4-p}\Big(\frac{\partial
\phi_{6-p}}{\partial y}\Big)^2 \Big] = (1-\nu_i^2)r^{7-p} + c_0^2
- \nu_i^2c_i^2  - r^2\omega_{3-p}^2\sin^2\phi_{2-p} \nonumber \\
&-& r^2\omega_{5-p}^2\cos^2\phi_{2-p}\sin^2\phi_{4-p} -
r^2\omega_{7-p}^2\cos^2\phi_{2-p}\cos^2\phi_{4-p}\sin^2\phi_{6-p}
\nonumber \\ &-&
r^2\omega_{8-p}^2\cos^2\phi_{2-p}\cos^2\phi_{4-p}\cos^2\phi_{6-p}
- \frac{\omega_{3-p}^2d_{3-p}^2}{r^{p-5}\sin^2\phi_{2-p}} -
\frac{\omega_{5-p}^2d_{5-p}^2}{r^{p-5}\cos^2\phi_{2-p}
\sin^2\phi_{4-p}} \nonumber \\ &-&
\frac{\omega_{7-p}^2d_{7-p}^2}{r^{p-5}\cos^2\phi_{2-p}\cos^2\phi_{4-p}
\sin^2\phi_{6-p}} - \frac{\omega_{8-p}^2d_{8-p}^2}{r^{p-5}
\cos^2\phi_{2-p}\cos^2\phi_{4-p} \cos^2\phi_{6-p}}  \ .
\label{Virasoro 3}
\end{eqnarray}
This Virasoro constraint (\ref{Virasoro 3}) suggests that we could
have some solutions, if we keep only one variable among $r,
\phi_{2-p}, \phi_{4-p}, \phi_{6-p}$ and rest of them to be
constants. But, putting some of the variables constants will
impose non trivial constraints on the system as said earlier. We
will analyse all those conditions one by one.

The conserved charges are given by,
\begin{eqnarray}
E = -\int \frac{\partial\mathcal{L}}{\partial \dot{t}} d\sigma &=&
\frac{\sqrt{k}}{2\pi(1-v^2)}\int[r^{\frac{7-p}{2}} - vc_0]d\sigma, \nonumber \\
P_i = \int \frac{\partial\mathcal{L}}{\partial \dot{x_i}} d\sigma &=&
\frac{\sqrt{k}\nu_i}{2\pi(1-v^2)}\int[r^{\frac{7-p}{2}} - vc_i]d\sigma, \nonumber \\
J_{\phi_{3-p}} = \int \frac{\partial\mathcal{L}}{\partial \dot{\phi}_{3-p}} d\sigma &=&
\frac{\sqrt{k}\omega_{3-p}}{2\pi(1-v^2)}\int[r^{\frac{p-3}{2}}\sin^2\phi_{2-p} - vd_{3-p}]d\sigma,
\nonumber \\  J_{\phi_{5-p}} = \int \frac{\partial\mathcal{L}}{\partial \dot{\phi}_{5-p}} d\sigma &=&
\frac{\sqrt{k}\omega_{5-p}}{2\pi(1-v^2)}\int[r^{\frac{p-3}{2}} \cos^2\phi_{2-p}\sin^2\phi_{4-p} -
vd_{5-p}]d\sigma, \nonumber \\  J_{\phi_{7-p}} = \int \frac{\partial\mathcal{L}}{\partial
\dot{\phi}_{7-p}} d\sigma &=& \frac{\sqrt{k}\omega_{7-p}}{2\pi(1-v^2)}\int[r^{\frac{p-3}{2}}
\cos^2\phi_{2-p}\cos^2\phi_{4-p}\sin^2\phi_{6-p} - vd_{7-p}]d\sigma, \nonumber \\  J_{\phi_{8-p}}
= \int \frac{\partial\mathcal{L}}{\partial \dot{\phi}_{8-p}} d\sigma &=& \frac{\sqrt{k}\omega_{8-p}}
{2\pi(1-v^2)}\int[r^{\frac{p-3}{2}} \cos^2\phi_{2-p}\cos^2\phi_{4-p}\cos^2\phi_{6-p} - vd_{8-p}]d\sigma \ .
\nonumber \\
\end{eqnarray}
Also the deficit angles are given by,
\begin{eqnarray}
\Delta \phi_{3-p} &=& \omega_{3-p}\int\frac{\partial g_{3-p}}{\partial y}d\sigma
= \frac{\omega_{3-p}}{1-v^2}\int\Big[ \frac{d_{3-p}}{r^{\frac{p-3}{2}}\sin^2\phi_{2-p}}
- v\Big]d\sigma, \nonumber \\ \Delta \phi_{5-p} &=& \omega_{5-p}\int\frac{\partial g_{5-p}}{\partial y}
d\sigma = \frac{\omega_{5-p}}{1-v^2}\int\Big[ \frac{d_{5-p}}{r^{\frac{p-3}{2}}
\cos^2\phi_{2-p}\sin^2\phi_{4-p}} - v\Big]d\sigma, \nonumber \\ \Delta \phi_{7-p}
&=& \omega_{7-p}\int\frac{\partial g_{7-p}}{\partial y}d\sigma =
\frac{\omega_{7-p}}{1-v^2}\int\Big[ \frac{d_{7-p}}{r^{\frac{p-3}{2}}
\cos^2\phi_{2-p}\cos^2\phi_{4-p}\sin^2\phi_{6-p}} - v\Big]d\sigma, \nonumber \\
\Delta \phi_{8-p} &=& \omega_{8-p}\int\frac{\partial g_{8-p}}{\partial y}d\sigma
= \frac{\omega_{8-p}}{1-v^2}\int\Big[ \frac{d_{8-p}}{r^{\frac{p-3}{2}}
\cos^2\phi_{2-p}\cos^2\phi_{4-p}\cos^2\phi_{6-p}} - v\Big]d\sigma \ . \nonumber \\
\end{eqnarray}
These relations can be reduced for the D1, D3 and D5- brane
backgrounds. In doing so, we should take the terms with only
positive subscript. In the next section, we will solve these
equations, corresponding to various rotating and orbiting string
ansatz explicitly for various D-brane backgrounds as mentioned
earlier.
\section{Rotating and Orbiting strings in D3-brane background}
Let's start our analysis from D3-brane, because the string
equations in the near horizon geometry of D3-brane are integrable.
In addition to the giant magnon and single spike solutions of the
string, we will further obtain more rotating and orbiting
solutions.
For $p=3$ equation (\ref{eom 2}) becomes,
\begin{eqnarray}
(1-v^2)^2\frac{\partial}{\partial
y}\Big[\cos^2\phi_1\frac{\partial \phi_3}{\partial y}\Big] &=&
\cos^2\phi_1\sin\phi_3\cos\phi_3
\Big[\frac{\omega_4^2d_4^2}{\cos^4\phi_1 \sin^4\phi_3} \nonumber \\ &-&
\frac{\omega_5^2d_5^2}{\cos^4\phi_1 \cos^4\phi_3} + \omega_5^2 -
\omega_4^2  \Big]
\label{eom 6}
\end{eqnarray}
and equation (\ref{eom 3}) becomes,
\begin{eqnarray}
&& (1-v^2)^2\frac{\partial}{\partial y}\Big[\frac{\partial
\phi_1}{\partial y}\Big] = \sin\phi_1\cos\phi_1 \Big[
\frac{\omega_2^2d_2^2}{\sin^4\phi_1} -
\frac{\omega_4^2d_4^2}{\cos^4\phi_1\sin^2\phi_3} \nonumber \\ &-&
\frac{\omega_5^2d_5^2}{\cos^4\phi_1\cos^2\phi_3} - \omega_2^2 +
\omega_4^2\sin^2\phi_3 + \omega_5^2\cos^2\phi_3 -
(1-v^2)^2\Big(\frac{\partial \phi_3}{\partial y}\Big)^2 \Big] \ .
\label{eom 7}
\end{eqnarray}
Equation (\ref{eom 5}) becomes,
\begin{eqnarray}
2(1-v^2)^2\frac{\partial}{\partial y} \Big[\frac{1}{r^2}
\frac{\partial r}{\partial y}\Big] = [1- \frac{c_0^2}{r^4} -
\nu_i^2\{1 - \frac{c_i^2}{r^4}\}]2r + (1-v^2)^2\Big(\frac{\partial
r}{\partial y}\Big)^2\partial_r(\frac{1}{r^2}) \ . \nonumber \\
\label{eom 8}
\end{eqnarray}
It can be noted from these equations of motion that the
differential equations corresponding to the radial variable ($r$)
and the angular variables $(\phi_1, \phi_3)$ are completely
decoupled. We will see later on that this very fact plays an
important role in getting exact solutions. Interestingly this
happens only for the D3 brane background only. Finally if
constraint (\ref{constraint 1}) is satisfied then the Virasoro
(\ref{Virasoro 1}) becomes,
\begin{eqnarray}
&& (1 - v^2)^2 \Big[\Big(\frac{\partial r}{\partial y}\Big)^2 +
r^2\Big(\frac{\partial \phi_1}{\partial y}\Big)^2 +
r^2\cos^2\phi_1\Big(\frac{\partial \phi_3}{\partial y}\Big)^2
\Big] = (1-\nu_i^2)r^{4} \nonumber \\ &+& c_0^2 - \nu_i^2c_i^2  -
r^2\omega_2^2\sin^2\phi_1 - r^2\omega_4^2\cos^2\phi_1\sin^2\phi_3
- r^2\omega_5^2\cos^2\phi_1\cos^2\phi_3 \nonumber \\ &-&
\frac{r^2\omega_2^2d_2^2}{\sin^2\phi_1} -
\frac{r^2\omega_4^2d_4^2}{\cos^2\phi_1 \sin^2\phi_3} -
\frac{r^2\omega_5^2d_5^2}{\cos^2\phi_1\cos^2\phi_3} \ .
\label{Virasoro 3}
\end{eqnarray}
As said earlier, it can be seen from this Virasoro (\ref{Virasoro
3}) that we would get some solutions if we keep only one variable
and put rest as constants. But as $r$ equation is decoupled from
$\phi_1$ and $\phi_3$ equations we can relax the above condition
as well and can have solutions if we keep $r$ and any one of
$\phi_1$ and  $\phi_3$ as variables.
In the following we will discuss different possible solutions.

\subsection{Rotating String Solutions: Constant ($r$, $\phi_3$) }
In this case $\phi_1$ is the only variable and $r$ and $\phi_3$
equations will generate constraints on the system. So, from
equation (\ref{eom 6}) we get the constraint,
\begin{equation}
\omega_4^2d_4^2\cos^4\phi_3 - \omega_5^2d_5^2\sin^4\phi_3 =
(\omega_4^2 - \omega_5^2)\cos^4\phi_1\sin^4\phi_3\cos^4\phi_3 \ .
\label{cnst 2}
\end{equation}
This constraint (\ref{cnst 2}) will imply $\phi_1 = {\rm
constant}$, which is our only variable. To avoid this we have to
put $\omega_4^2 = \omega_5^4$. Under this condition the constraint
(\ref{cnst 2}) becomes,
\begin{equation}
d_4^2\cos^4\phi_3 = d_5^2\sin^4\phi_3 \ . \label{cnst 3}
\end{equation}
Using the above constraint (\ref{cnst 3}), we get, from equation
(\ref{eom 7}),
\begin{eqnarray}
&& (1-v^2)^2\frac{\partial^2 \phi_1}{\partial y^2} =
\sin\phi_1\cos\phi_1 \Big[ \frac{\omega_2^2d_2^2}{\sin^4\phi_1} -
\frac{\omega_4^2d_4^2}{\cos^4\phi_1\sin^4\phi_3}  - \omega_2^2 +
\omega_4^2 \Big] \ . \label{eom 9}
\end{eqnarray}
Integrating equation (\ref{eom 9}), we have,
\begin{equation}
(1-v^2)^2\Big(\frac{\partial \phi_1}{\partial y}\Big)^2 =
-\frac{\omega_2^2d_2^2}{\sin^2\phi_1} -
\frac{\omega_4^2d_4^2}{\cos^2\phi_1\sin^4\phi_3} - (\omega_2^2 -
\omega_4^2)\sin^2\phi_1 + c_4 \ . \label{eom 10}
\end{equation}
Equation (\ref{eom 8}) will give the constraint,
\begin{equation}
(1-\nu_i^2)r^4 = c_0^2 - \nu_i^2c_i^2 \ . \label{cnst 4}
\end{equation}
Thus the constraints (\ref{cnst 3}) and (\ref{cnst 4}) fixes the values of the constants $\phi_3$ and $r$ respectively.
Finally using all the constraints we get from the Virasoro (\ref{Virasoro 3}),
\begin{eqnarray}
(1-v^2)^2\Big(\frac{\partial \phi_1}{\partial y}\Big)^2 &=&
-\frac{\omega_2^2d_2^2}{\sin^2\phi_1} -
\frac{\omega_4^2d_4^2}{\cos^2\phi_1\sin^4\phi_3} \nonumber \\ &-&
\omega_2^2\sin^2\phi_1 - \omega_4^2\cos^2\phi_1 + \frac{2(c_0^2 -
\nu_i^2c_i^2)}{r^2} \ . \label{Virasoro 4}
\end{eqnarray}
Comparing equation (\ref{eom 10}) and equation (\ref{Virasoro 4}) we get,
\begin{equation}
c_4 = \frac{2(c_0^2 - \nu_i^2c_i^2)}{r^2} - \omega_2^2 \ .
\label{cnst 5}
\end{equation}
Using the limit $\frac{\partial\phi_1}{\partial y} \rightarrow 0$
as $\phi_1 \rightarrow \frac{\pi}{2}$ in equation (\ref{eom 10})
we get, $d_4 = 0$ and $c_4 = \omega_2^2d_2^2 + \omega_2^2 -
\omega_4^2$. Using this equation (\ref{eom 10}) becomes,
\begin{equation}
\frac{\partial \phi_1}{\partial y} = \frac{\sqrt{\omega_2^2 -
\omega_4^2}}{1-v^2} \cot\phi_1 \sqrt{\sin^2\phi_1 -
\sin^2\phi_{min}} \ ,
\end{equation}
where $\sin\phi_{min} = \frac{\omega_2d_2}{\sqrt{\omega_2^2 -
\omega_4^2}}$. Below we will discuss two different cases
corresponding to giant magnon and single spike solutions for the
string separately,

\subsubsection{Giant magnon Case}
For $d_2 = v$,
\begin{equation}
\Delta\phi_2 = 2\arccos(\sin\phi_{min}) \Rightarrow \sin\phi_{min}
= \cos \frac{\Delta\phi}{2} \ ,
\end{equation}
where we have taken $\Delta\phi = \Delta\phi_2$. It is easy to see that $E$, $P_i$ and $J_{\phi_2}$ are divergent
independently, but the combination,
\begin{equation}
\tilde{E} - J_{\phi_2} =
\frac{\sqrt{k}\omega_2}{\pi\sqrt{\omega_2^2 - \omega_4^2}}
\cos\phi_{min} \ , \label{rel 01}
\end{equation}
is finite, where $\tilde{E} = \frac{(1-v^2)\omega_2}{v(c_i - c_0)}
(E - \frac{P_i}{\nu_i})$. $J_{\phi_4}$ and $J_{\phi_5}$ are finite
and given by,
\begin{eqnarray}
J_{\phi_4} &=& \frac{\sqrt{k}\omega_4}{\pi \sqrt{\omega_2^2 -
\omega_4^2}} \sin^2\phi_3 \cos\phi_{min} , \>\>\> J_{\phi_5} =
\frac{\sqrt{k}\omega_5}{\pi \sqrt{\omega_2^2 - \omega_4^2}}
\cos^2\phi_3 \cos\phi_{min} \ .
\end{eqnarray}
Let us define,
\begin{equation}
J_{\phi} = \omega \Big(\frac{J_{\phi_4}}{\omega_4} +
\frac{J_{\phi_5}}{\omega_5}\Big) = \frac{\sqrt{k}\omega}{\pi
\sqrt{\omega_2^2 - \omega_4^2}} \cos\phi_{min} \ . \label{rel 02}
\end{equation}
Combining equation (\ref{rel 01}) and equation (\ref{rel 02}), we
get the dyonic like giant magnon dispersion relation,
\begin{equation}
\tilde{E} - J_{\phi_2} = \sqrt{J_{\phi}^2 + \frac{k(\omega_2^2 -
\omega^2)}{\pi^2(\omega_2^2 - \omega_4^2)} \sin^2
\Big(\frac{\Delta\phi}{2}\Big)} \ . \label{magnon}
\end{equation}
\subsubsection{Single Spike Case}
For $d_2 = \frac{1}{v}$, $\Delta\phi_2$, $E$ and $P_i$ are divergent, but the combination,
\begin{equation}
(\Delta\phi_2)_{reg} = \Delta\phi_2 -
\frac{(1-v^2)2\pi\omega_2d_2}{\sqrt{k}v(c_i-c_0)} \Big(E -
\frac{P_i}{\nu_i}\Big) = -2\arccos(\sin\phi_{min}) \ ,
\end{equation}
is finite, which implies $\sin\phi_{min} = \cos
\frac{(\Delta\phi)_{reg}}{2}$, where we rename $(\Delta\phi_2)_{reg} = (\Delta\phi)_{reg}$. All the angular momenta are finite
here,
\begin{eqnarray}
J_{\phi_2} &=& \frac{\sqrt{k}\omega_2}{\pi \sqrt{\omega_2^2 -
\omega_4^2}} \cos\phi_{min} , \nonumber \\  J_{\phi_4} &=&
-\frac{\sqrt{k}\omega_4}{\pi \sqrt{\omega_2^2 - \omega_4^2}}
\sin^2\phi_3 \cos\phi_{min} , \nonumber \\ J_{\phi_5} &=&
-\frac{\sqrt{k}\omega_5}{\pi \sqrt{\omega_2^2 - \omega_4^2}}
\cos^2\phi_3 \cos\phi_{min}\ .
\end{eqnarray}
Again defining the combination,
\begin{equation}
J_{\phi} = \omega \Big(\frac{J_{\phi_4}}{\omega_4} +
\frac{J_{\phi_5}}{\omega_5}\Big) = -\frac{\sqrt{k}\omega}{\pi
\sqrt{\omega_2^2 - \omega_4^2}} \cos\phi_{min} \ .
\end{equation}
Combining these relations we can write the dispersion relation,
\begin{equation}
J_{\phi_2} = \sqrt{J_{\phi}^2 + \frac{k(\omega_2^2 -
\omega^2)}{\pi^2(\omega_2^2 - \omega_4^2)} \sin^2
\Big(\frac{(\Delta\phi)_{reg}}{2}\Big)} \ , \label{spike}
\end{equation}
which is nothing but the well known spike dispersion relation.
However, there are further rigidly rotating string solutions on
this subspace of solutions which we discuss below.
\subsection{Rotating String Solution: Constant ($r$, $\phi_1$)}
In this case $\phi_3$ is the only variable. We have from equation (\ref{eom 6}),
\begin{equation}
(1-v^2)^2\frac{\partial}{\partial y}\Big[\frac{\partial
\phi_3}{\partial y}\Big] = \sin\phi_3\cos\phi_3
\Big[\frac{\omega_4^2d_4^2}{\cos^4\phi_1 \sin^4\phi_3} -
\frac{\omega_5^2d_5^2}{\cos^4\phi_1 \cos^4\phi_3} + \omega_5^2 -
\omega_4^2  \Big] \ , \label{eom 11}
\end{equation}
Integrating equation (\ref{eom 11}) we get,
\begin{equation}
(1-v^2)^2\Big(\frac{\partial \phi_3}{\partial y}\Big)^2 =
-\frac{1}{\cos^4\phi_1} \Big[\frac{\omega_4^2d_4^2}{ \sin^2\phi_3}
+ \frac{\omega_5^2d_5^2}{ \cos^2\phi_3}\Big] + (\omega_5^2 -
\omega_4^2)\sin^2\phi_3 + c_4   \ . \label{eom 12}
\end{equation}
Again equation (\ref{eom 7}) becomes,
\begin{eqnarray}
(1-v^2)^2\Big(\frac{\partial \phi_3}{\partial y}\Big)^2 &=&
-\frac{1}{\cos^4\phi_1} \Big[\frac{\omega_4^2d_4^2}{ \sin^2\phi_3}
+ \frac{\omega_5^2d_5^2}{ \cos^2\phi_3}\Big] +
\frac{\omega_2^2d_2^2}{\sin^4\phi_1} \nonumber \\ &-& \omega_2^2 +
\omega_4^2\sin^2\phi_3 + \omega_5^2\cos^2\phi_3  \ . \label{eom
13}
\end{eqnarray}
Comparing equation (\ref{eom 12}) and (\ref{eom 13}) we get,
\begin{equation}
c_4 = 2(\omega_4^2 - \omega_5^2)\sin^2\phi_3 - \omega_2^2 +
\omega_5^2 + \frac{\omega_2^2d_2^2}{\sin^4\phi_1} \ . \label{cnst
6}
\end{equation}
This constraint (\ref{cnst 6}) implies $\phi_3$ is constant, which
is our only variable here. To avoid this we must put $\omega_4^2 =
\omega_5^2$. Using this the constraint (\ref{cnst 6}) reduces to,
\begin{equation}
c_4 =  - \omega_2^2 + \omega_5^2 +
\frac{\omega_2^2d_2^2}{\sin^4\phi_1} \ , \label{cnst 7}
\end{equation}
and equation (\ref{eom 12}) becomes,
\begin{equation}
(1-v^2)^2\Big(\frac{\partial \phi_3}{\partial y}\Big)^2 =
-\frac{\omega_4^2}{\cos^4\phi_1} \Big[\frac{d_4^2}{ \sin^2\phi_3}
+ \frac{d_5^2}{ \cos^2\phi_3}\Big] + c_4   \ . \label{eom 14}
\end{equation}
Equation (\ref{eom 8}) gives the constraint,
\begin{equation}
(1-\nu_i^2)r^4=c_0^2 - \nu_i^2c_i^2 \ . \label{cnst 8}
\end{equation}
and finally using all the constraints in  Virasoro (\ref{Virasoro 3}) we get,
\begin{eqnarray}
&& (1-v^2)^2\Big(\frac{\partial \phi_3}{\partial y}\Big)^2 =
-\frac{\omega_4^2}{\cos^4\phi_1} \Big[\frac{d_4^2}{ \sin^2\phi_3}
+ \frac{d_5^2}{ \cos^2\phi_3}\Big] - \omega_4^2 \nonumber \\ &+&
\frac{1}{\cos^2\phi_1} \Big[\frac{2(c_0^2 - \nu_i^2c_i^2)}{r^2} -
\omega_2^2\sin^2\phi_1 - \frac{\omega_2^2d_2^2}{\sin^2\phi_1}\Big]
\ . \label{Virasoro 5}
\end{eqnarray}
Again comparing equation (\ref{eom 14}) with equation (\ref{Virasoro 5}) we get,
\begin{equation}
c_4 = \frac{1}{\cos^2\phi_1} \Big[\frac{2(c_0^2 -
\nu_i^2c_i^2)}{r^2} - \omega_2^2\sin^2\phi_1 -
\frac{\omega_2^2d_2^2}{\sin^2\phi_1}\Big] - \omega_4^2 \ .
\label{cnst 9}
\end{equation}
Finally comparing equation (\ref{cnst 8}) with equation (\ref{cnst 9}) we get,
\begin{equation}
2\omega_2^2\sin^2\phi_1 - \omega_2^2 + 2\omega_4^2\cos^2\phi_1 +
\frac{\omega_2^2d_2^2}{\sin^4\phi_1} = 2\sqrt{(1-\nu_i^2)(c_0^2 -
\nu_i^2c_i^2)} \ . \label{cnst 10}
\end{equation}
Using all these constraints finally we get,
\begin{equation}
\frac{\partial \phi_3}{\partial y} = \frac{\sqrt{c_4}}{1-v^2}
\frac{\sqrt{(\sin^2\phi_3 - \sin^2\phi_{min})(\sin^2\phi_{max} -
\sin^2\phi_3)}}{\sin\phi_3 \cos\phi_3} \ ,
\end{equation}
where $\sin^2\phi_{max} + \sin^2\phi_{min} = 1 +
\frac{\omega_4^2(d_4^2 - d_5^2)}{c_4^2\cos^4\phi_1}$ and
$\sin^2\phi_{max} \sin^2\phi_{min} =
\frac{\omega_4^2d_4^2}{c_4^2\cos^4\phi_1}$. In this case, all the
conserved charges are finite when we integrate $\phi_3$ from $\phi_{min}$ to $\phi_{max}$, the explicit expression for these
are given by,
\begin{eqnarray}
E &=& \pm \frac{\sqrt{k}(r^2-vc_0)}{2\sqrt{c_4}} \ , \>\>\> P_i =
\pm \frac{\sqrt{k}\nu_i(r^2-vc_i)}{2\sqrt{c_4}} \ ,
\end{eqnarray}
we can combine them as
\begin{equation}
E - \frac{1}{3}\sum_{i=1,2,3}\frac{P_i}{\nu_i} = \pm \frac{\sqrt{k}v}{2\sqrt{c_4}}\Big[\frac{1}{3}\sum_i c_i - c_0\Big] \ . \label{rel 8}
\end{equation}
Also,
\begin{eqnarray}
J_{\phi_2} &=& \pm \frac{\sqrt{k}\omega_2}{2\sqrt{c_4}} [\sin^2\phi_1 - vd_2],
\nonumber \\ J_{\phi_4} &=& \pm \frac{\sqrt{k}\omega_4}{4\sqrt{c_4}} [\cos^2\phi_1
(\sin^2\phi_{min} + \sin^2\phi_{max}) - 2vd_4], \nonumber \\
J_{\phi_5} &=& \pm \frac{\sqrt{k}\omega_5}{4\sqrt{c_4}} [
2(\cos^2\phi_1 - vd_5) - \cos^2\phi_1(\sin^2\phi_{min} +
\sin^2\phi_{max})] \ .
\end{eqnarray}
We can also combine,
\begin{eqnarray}
&& \frac{J_{\phi_2}}{\omega_2} + \frac{J_{\phi_4}}{\omega_4} +
\frac{J_{\phi_5}}{\omega_5}  = \pm \frac{\sqrt{k}}{2\sqrt{c_4}}[1
- v( d_2 + d_4 + d_5)] \ . \label{rel 9}
\end{eqnarray}
Combining equation (\ref{rel 8}) and equation (\ref{rel 9}), we get,
\begin{equation}
E - \frac{1}{3}\sum_{i=1,2,3}\frac{P_i}{\nu_i} -\sum_{j=2,4,5}\frac{J_{\phi_j}}{\omega_j} =
\pm\frac{\sqrt{k}}{2\sqrt{c_4}} \Big[v(-c_0 + \frac{1}{3}\sum_ic_i + \sum_jd_j) -
1\Big] \ . \label{rot 1}
\end{equation}
We can't relate this equation with the deficit angles, so we are
not writing their expressions explicitly. This is a new kind of rotating
string solution which is of the form $E - J =$ constant, for constant $E$ and $J$. One could think them as dual to
some chiral primary operators on the gauge theory side, however we
would like to mention that the ansatz that we have proposed here are not
the typical closed string ansatz, rather they are for the open
strings.
\subsection{Orbiting String Solution: Constant ($\phi_1$, $\phi_3$)}
In this case $r$ is the only variable. From equation (\ref{eom 6}) we have
\begin{equation}
\omega_4^2d_4^2\cos^4\phi_3 - \omega_5^2d_5^2\sin^4\phi_3 =
(\omega_4^2 - \omega_5^2)\cos^4\phi_1\sin^4\phi_3\cos^4\phi_3 \ .
\label{const 01}
\end{equation}
Also from equation (\ref{eom 7}) we have,
\begin{eqnarray}
&& \omega_2^2d_2^2\sin^2\phi_3\cos^2\phi_3 -
\sin^4\phi_1(\omega_4^2d_4^2\cos^2\phi_3 +
\omega_5^2d_5^2\sin^2\phi_3) \nonumber \\ &=& (\omega_2^2 -
\omega_4^2\sin^2\phi_3 - \omega_5^2\cos^2\phi_3)
\sin^4\phi_1\cos^4\phi_1\sin^2\phi_3\cos^2\phi_3 \ . \label{cnst
02}
\end{eqnarray}
From Virasoro (\ref{Virasoro 3}) we have,
\begin{equation}
(1-v^2)^2\Big(\frac{\partial r}{\partial y}\Big)^2 = (1 -
\nu_i^2)r^4 - c_4r^2 + (c_0^2 - \nu_i^2c_i^2) \ , \label{eom 01}
\end{equation}
where $c_4 = \omega_2^2\sin^2\phi_1 +
\omega_4^2\cos^2\phi_1\sin^2\phi_3 +
\omega_5^2\cos^2\phi_1\cos^2\phi_3 +
\frac{\omega_2^2d_2^2}{\sin^2\phi_1} +
\frac{\omega_4^2d_4^2}{\cos^2\phi_1\sin^2\phi_3} +
\frac{\omega_5^2d_5^2}{\cos^2\phi_1\cos^2\phi_3}$. Substituting
the value of $\frac{\partial r}{\partial y}$ from (\ref{eom 01})
into equation (\ref{eom 8}) we get,
\begin{equation}
2(1-v^2)^2 \frac{\partial^2 r}{\partial y^2} = 4(1-\nu_i^2)r^3 -
2c_4r \ . \label{eom 02}
\end{equation}
Integrating equation (\ref{eom 02}) we get,
\begin{equation}
(1-v^2)^2\Big(\frac{\partial r}{\partial y}\Big)^2 = (1 -
\nu_i^2)r^4 - c_4r^2 + c_5 \ . \label{eom 03}
\end{equation}
Comparing equation (\ref{eom 01}) with equation (\ref{eom 03}) we get,
\begin{equation}
c_5 = c_0^2 - \nu_i^2c_i^2 \ .
\end{equation}
So, finally we have
\begin{equation}
\frac{\partial r}{\partial y} = \frac{\sqrt{1 - \nu_i^2}}{1-v^2}
\sqrt{(r^2 - a^2)(r^2 - b^2)} \ ,
\end{equation}
where $a^2 + b^2 = \frac{c_4}{1 - \nu_i^2}$ and $a^2b^2 = \frac{c_5}{1 - \nu_i^2}$.
Now the conserved charges in this case are given by,
\begin{eqnarray}
E &=& \frac{\sqrt{k}}{\pi \sqrt{1 - \nu_i^2}} [I_1 -vc_0I_2] ,
\nonumber \\  P_i &=& \frac{\sqrt{k}\nu_i}{\pi \sqrt{1 - \nu_i^2}}
[I_1 -vc_iI_2] , \nonumber \\  J_{\phi_2} &=& \frac{\sqrt{k}
\omega_2}{\pi \sqrt{1 - \nu_i^2}} (\sin^2\phi_1 - vd_2)I_2 ,
\nonumber \\ J_{\phi_4} &=& \frac{\sqrt{k} \omega_4}{\pi \sqrt{1 -
\nu_i^2}} (\cos^2\phi_1\sin^2\phi_3 - vd_4)I_2 , \nonumber \\
J_{\phi_5} &=& \frac{\sqrt{k} \omega_5}{\pi \sqrt{1 - \nu_i^2}}
(\cos^2\phi_1\cos^2\phi_3 - vd_5)I_2 \ ,
\end{eqnarray}
where
\begin{eqnarray}
I_1 = \int \frac{r^2 dr}{\sqrt{(r^2 - a^2)(r^2 - b^2)}} &=& \pm b
\Big[E[i\sinh^{-1}\Big(\frac{ir}{a}\Big), \frac{a^2}{b^2}]
\nonumber \\ &-& F[i\sinh^{-1}\Big(\frac{ir}{a}\Big),
\frac{a^2}{b^2}]\Big], \nonumber \\ I_2 = \int
\frac{dr}{\sqrt{(r^2 - a^2)(r^2 - b^2)}} &=& \pm \frac{1}{b}
F[\sin^{-1}\Big(\frac{r}{a}\Big), \frac{a^2}{b^2}] \ . \label{int}
\end{eqnarray}
where $F(\varphi,m)$ and $E(\varphi,m)$ are the incomplete
elliptic integrals of first and second kind respectively. Note that in this case we have not use any integration limit. Now we
can combine different charges to find a relationship among
themselves, firstly combining $E$ and $P_i$ we get,
\begin{equation}
E - \frac{P_i}{\nu_i} = \frac{\sqrt{k}}{\pi\sqrt{1-\nu_i^2}}v(c_i
- c_0)I_2 \ ,
\end{equation}
then combining the $J_{\phi}$'s we get,
\begin{equation}
\sum_{j=2,4,5}\frac{J_{\phi_j}}{\omega_j} = \frac{J_{\phi_2}}{\omega_2} + \frac{J_{\phi_4}}{\omega_4} +
\frac{J_{\phi_5}}{\omega_5} =
\frac{\sqrt{k}}{\pi\sqrt{1-\nu_i^2}}[1 - v\sum_j d_j]I_2 \
,
\end{equation}
and combining these two relations we get,
\begin{equation}
\sum_{j=2,4,5}\frac{J_{\phi_j}}{\omega_j}  = \frac{ 1 - v\sum_j
d_j}{v(c_i - c_0)} \Big(E - \frac{P_i}{\nu_i}\Big)  \ . \label{orb
1}
\end{equation}
These are some new solutions which are obtained from a particular embedding of worldsheet variables of the string in the stack of
D3-brane background. To have a better understanding of these solutions we can use a particular type of integration limit. For example, if we use the limit $\frac{\partial r}{\partial y} \to 0$ as $r \to 0$, then from equation (\ref{eom 03}) we immediately get $c_5=0$ and it becomes,
\begin{equation}
    \frac{\partial r}{\partial y} = \frac{\sqrt{1-\nu_i^2}}{1-v^2}r\sqrt{r^2 - r_0^2} \ ,
\end{equation}
where $r_0 = \frac{c_4}{1-\nu_i^2}$. In this case we can integrate $r$ from $0$ to $r_0$ and find from (\ref{int}) that $I_1$ is finite, while $I_2$ diverges. This implies all the conserved charges diverge including the deficit angles. However, the relation (\ref{orb 1}) is still true and the solution have the form $E - J = 0$ for diverging $E$ and $J$. It will be nice to be able to tell about the corresponding operators in detail.
\section{Strings in D5-brane background}
In this section we wish to study the rotating and orbiting strings in the
D5-brane background. For $p=5$ equation (\ref{eom 2}) becomes,
\begin{equation}
(1-v^2)^2\frac{\partial}{\partial y}\Big[r\frac{\partial
\phi_1}{\partial y}\Big] = r\sin\phi_1\cos\phi_1
\Big[\frac{\omega_2^2d_2^2}{r^2 \sin^4\phi_1} -
\frac{\omega_3^2d_3^2}{r^2 \cos^4\phi_1} + \omega_3^2 - \omega_2^2
\Big] \ , \label{eom 15}
\end{equation}
and equation (\ref{eom 5}) becomes,
\begin{eqnarray}
&& 2(1-v^2)^2\frac{\partial}{\partial y} \Big[\frac{1}{r}
\frac{\partial r}{\partial y}\Big] = (1-\nu_i^2) - \frac{c_0^2 -
\nu_i^2c_i^2}{r^2} + (1-v^2)^2\Big(\frac{\partial r}{\partial
y}\Big)^2\partial_r(\frac{1}{r}) \nonumber \\ &+&
\frac{\omega_2^2d_2^2}{r^2\sin^2\phi_1}
+\frac{\omega_3^2d_3^2}{r^2\cos^2\phi_1} - \omega_2^2\sin^2\phi_1
- \omega_3^2\cos^2\phi_1 + (1-v)^2\Big(\frac{\partial
\phi_1}{\partial y}\Big)^2 \ .  \label{eom 16}
\end{eqnarray}
As one can note that these system of equations are coupled and one
can not solve for $r$ and $\phi_1$ independently as in the case of
D3 brane background. Finally if constraint (\ref{constraint 1}) is
satisfied then the Virasoro (\ref{Virasoro 1}) becomes,
\begin{eqnarray}
&& (1 - v^2)^2 \Big[\Big(\frac{\partial r}{\partial y}\Big)^2 +
r^2\Big(\frac{\partial \phi_1}{\partial y}\Big)^2 \Big] =
(1-\nu_i^2)r^{2} + c_0^2 - \nu_i^2c_i^2  \nonumber \\ &-&
r^2\omega_2^2\sin^2\phi_1 - r^2\omega_3^2\cos^2\phi_1 -
\frac{\omega_2^2d_2^2}{\sin^2\phi_1} -
\frac{\omega_3^2d_3^2}{\cos^2\phi_1} \ . \label{Virasoro 6}
\end{eqnarray}
Now by looking at the Virasoro (\ref{Virasoro 6}), to get some
solution we have to put one of the variable to be constant. In the
following we will discuss the different possible solutions.
\subsection{Rotating String Solution: Constant ($r$)}
Here $\phi_1$ is the only variable, and equation (\ref{eom 15})
becomes,
\begin{equation}
(1-v^2)^2\frac{\partial}{\partial y}\Big[\frac{\partial
\phi_1}{\partial y}\Big] = \sin\phi_1\cos\phi_1
\Big[\frac{\omega_2^2d_2^2}{r^2 \sin^4\phi_1} -
\frac{\omega_3^2d_3^2}{r^2 \cos^4\phi_1} + \omega_3^2 - \omega_2^2
\Big] \ , \label{eom 17}
\end{equation}
Integrating (\ref{eom 17}) we get,
\begin{equation}
(1-v^2)^2\Big(\frac{\partial \phi_1}{\partial y}\Big)^2 =
-\frac{1}{r^2} \Big[\frac{\omega_2^2d_2^2}{ \sin^2\phi_1} +
\frac{\omega_3^2d_3^2}{ \cos^2\phi_1} \Big] + (\omega_3^2 -
\omega_2^2)\sin^2\phi_1 + c_6 \ , \label{eom 18}
\end{equation}
Now from equation (\ref{eom 16}) we have,
\begin{eqnarray}
(1-v^2)^2\Big(\frac{\partial \phi_1}{\partial y}\Big)^2 &=&
-\frac{1}{r^2} \Big[\frac{\omega_2^2d_2^2}{ \sin^2\phi_1} +
\frac{\omega_3^2d_3^2}{ \cos^2\phi_1} \Big] +
\omega_2^2\sin^2\phi_1 + \omega_3^2\cos^2\phi_1 \nonumber \\ &-&
(1-\nu_i^2) + \frac{c_0^2 - \nu_i^2c_i^2}{r^2} \ . \label{eom 19}
\end{eqnarray}
Comparing equation (\ref{eom 18}) with equation (\ref{eom 19}) we get,
\begin{equation}
c_6 = 2(\omega_2^2 - \omega_3^2)\sin^2\phi_1 + \omega_3^2 -
(1-\nu_i^2) + \frac{c_0^2 - \nu_i^2c_i^2}{r^2} \ . \label{cnst 11}
\end{equation}
This constraint (\ref{cnst 11}) implies $\phi_1$ is constant which is our only variable. To aviod this we should put $\omega_2^2 = \omega_3^2$. Under this condition the constraint (\ref{cnst 11}) reduces to,
\begin{equation}
c_6 =\omega_2^2 - (1-\nu_i^2) + \frac{c_0^2 - \nu_i^2c_i^2}{r^2} \
, \label{cnst 12}
\end{equation}
and equation (\ref{eom 18}) reduces to,
\begin{equation}
(1-v^2)^2\Big(\frac{\partial \phi_1}{\partial y}\Big)^2 =
-\frac{\omega_2^2}{r^2} \Big[\frac{d_2^2}{ \sin^2\phi_1} +
\frac{d_3^2}{ \cos^2\phi_1} \Big] + c_6 \ . \label{eom 20}
\end{equation}
Finally from Virasoro we have,
\begin{equation}
(1-v^2)^2\Big(\frac{\partial \phi_1}{\partial y}\Big)^2 =
-\frac{\omega_2^2}{r^2} \Big[\frac{d_2^2}{ \sin^2\phi_1} +
\frac{d_3^2}{ \cos^2\phi_1} \Big] - \omega_2^2 + (1-\nu_i^2) +
\frac{c_0^2 - \nu_i^2c_i^2}{r^2} \ . \label{Virasoro 7}
\end{equation}
Again comparing equation (\ref{eom 20}) with equation (\ref{Virasoro 7}) we get,
\begin{equation}
c_6 = (1-\nu_i^2) + \frac{c_0^2 - \nu_i^2c_i^2}{r^2} - \omega_2^2
\ , \label{cnst 13}
\end{equation}
Comparing these two $c_6$'s we get the constraint,
\begin{eqnarray}
\omega_2^2 &=& (1 - \nu_i^2), \>\>\> c_6 = \frac{c_0^2 -
\nu_i^2c_i^2}{r^2} \ . \label{cnst 14}
\end{eqnarray}
Using all these conditions and constraints we get,
\begin{equation}
\frac{\partial \phi_1}{\partial y} = \frac{\sqrt{c_6}}{1-v^2}
\frac{\sqrt{(\sin^2\phi_1 - \sin^2\phi_{min})(\sin^2\phi_{max} -
\sin^2\phi_1)}}{\sin\phi_1\cos\phi_1} \ ,
\end{equation}
where $\sin^2\phi_{min} + \sin^2\phi_{max} = 1 + \frac{\omega_2^2(d_2^2 - d_3^2)}{c_6r^2}$ and $\sin^2\phi_{min} \sin^2\phi_{max} = \frac{\omega_2^2d_2^2}{c_6r^2}$.
In this case, all the conserved charges are finite when we integrate $\phi_1$ from $\phi_{min}$ to $\phi_{max}$, the explicit expression for these are given by,
\begin{eqnarray}
E &=& \pm \frac{\sqrt{k}(r^2-vc_0)}{2\sqrt{c_6}} \ , \>\>\> P_i =
\pm \frac{\sqrt{k}\nu_i(r^2-vc_i)}{2\sqrt{c_6}} \ ,
\end{eqnarray}
we can combine them as
\begin{equation}
E - \frac{1}{5}\sum_{i=1}^5\frac{P_i}{\nu_i} = \pm \frac{\sqrt{k}v}{2\sqrt{c_6}} \Big[\frac{1}{5}\sum_i c_i - c_0\Big] \ . \label{rel 10}
\end{equation}
 Also,
\begin{eqnarray}
J_{\phi_2} &=& \pm \frac{\sqrt{k}\omega_2}{4\sqrt{c_6}} [r(\sin^2\phi_{min}
+ \sin^2\phi_{max}) - 2vd_2], \nonumber \\
J_{\phi_3} &=& \pm \frac{\sqrt{k}\omega_3}{4\sqrt{c_6}} [ 2(r -
vd_3) - r(\sin^2\phi_{min} + \sin^2\phi_{max})] \ .
\end{eqnarray}
We can also combine,
\begin{eqnarray}
&& \frac{J_{\phi_2}}{\omega_2} + \frac{J_{\phi_3}}{\omega_3}   =
\pm \frac{\sqrt{k}}{2\sqrt{c_6}}[r - v(d_2 + d_3)] \ . \label{rel
11}
\end{eqnarray}
Combining equation (\ref{rel 10}) and equation (\ref{rel 11}), we get,
\begin{equation}
E - \frac{1}{5}\sum_{i=1}^5\frac{P_i}{\nu_i} -\sum_{j=2,3}\frac{J_{\phi_j}}{\omega_j} =
\pm\frac{\sqrt{k}}{2\sqrt{c_6}} \Big[v(-c_0 + \frac{1}{5}\sum_ic_i + \sum_jd_j) - r\Big] \
. \label{rot 2}
\end{equation}
As said in the previous section, we can't relate this equation
with the deficit angles, so we are not writing their expressions
explicitly. Since $r$ is constant in this case, the right hand
side of the final equation (\ref{rot 2}) is again constant and
this relation is of the type $E - J$ is constant, for $E$ constant
and $J$ constant, as we obtain for the case of D3 branes equation
(\ref{rot 1}).

\subsection{Orbiting String Solution: Constant ($\phi_1$)}
In this case $r$ is the only variable. So, from equation (\ref{eom 15}) we have,
\begin{equation}
\omega_2^2d_2^2\cos^4\phi_1 - \omega_3^2d_3^2\sin^4\phi_1 =
(\omega_2^2 - \omega_3^2)r^2\sin^4\phi_1 \ . \label{cnst 15}
\end{equation}
This constraint (\ref{cnst 15}) implies $r$ is constant which is
our only variable. To avoid this we must put $\omega_2^2 =
\omega_3^2$. Under this condition the constraint (\ref{cnst 15})
reduces to,
\begin{equation}
d_2^2\cos^4\phi_1 =d_3^2\sin^4\phi_1 \ . \label{cnst 16}
\end{equation}
Now from equation (\ref{eom 16}) we have,
\begin{eqnarray}
2(1-v^2)^2\frac{1}{r}\frac{\partial^2 r}{\partial y^2} &=&
(1-v^2)^2 \frac{1}{r^2}\Big(\frac{\partial r}{\partial y}\Big)^2 +
(1 - \nu_i^2) \nonumber \\ &-& \frac{c_0^2 - \nu_i^2c_i^2}{r^2} +
\frac{\omega_2^2d_2^2}{r^2\sin^4\phi_1} - \omega_2^2 \ .
\label{eom 21}
\end{eqnarray}
Again from Virasoro (\ref{Virasoro 6}) we have,
\begin{equation}
(1-v^2)^2\Big(\frac{\partial r}{\partial y}\Big)^2 = (1 -
\nu_i^2)r^2 + (c_0^2 - \nu_i^2c_i^2) - r^2\omega_2^2 -
\frac{\omega_2^2d_2^2}{\sin^4\phi_1}  . \label{Virasoro 8}
\end{equation}
Substituting the value of $(\frac{\partial r}{\partial y})^2$ from
Virasoro (\ref{Virasoro 8}) into equation (\ref{eom 21}) we get,
\begin{equation}
(1-v^2)^2\frac{\partial^2r}{\partial y^2} = (1-\nu_i^2 -
\omega_2^2)r \ . \label{eom 22}
\end{equation}
Integrating (\ref{eom 22}) we get,
\begin{equation}
(1-v^2)^2\Big(\frac{\partial r}{\partial y}\Big)^2 = (1-\nu_i^2 -
\omega_2^2)r^2 + c_6 \ . \label{eom 23}
\end{equation}
Comparing equation (\ref{Virasoro 8}) with equation (\ref{eom 23}) we get,
\begin{equation}
c_6 = c_0^2 - \nu_i^2c_i^2 - \frac{\omega_2^2d_2^2}{\sin^4\phi_1}
\ . \label{cnst 17}
\end{equation}
Therefore we have,
\begin{equation}
\frac{\partial r}{\partial y} = \frac{\sqrt{1 - \nu_i^2 -
\omega_2^2}}{1-v^2} \sqrt{r^2 + r_0^2} \ ,
\end{equation}
where $r_0^2 = \frac{c_6}{1 - \nu_i^2 - \omega_2^2}$.
In this case the conserved quantities becomes,
\begin{eqnarray}
E &=& \frac{\sqrt{k}}{\pi\sqrt{1-\nu_i^2 - \omega_2^2}} [I_1 -
vc_0 I_2 ], \nonumber \\ P_i &=&
\frac{\sqrt{k}\nu_i}{\pi\sqrt{1-\nu_i^2 - \omega_2^2}} [I_1 - vc_i
I_2], \nonumber \\  J_{\phi_2} &=&
\frac{\sqrt{k}\omega_2}{\pi\sqrt{1-\nu_i^2 - \omega_2^2}}
[\sin^2\phi_1 I_1 - vd_2 I_2], \nonumber \\ J_{\phi_3} &=&
\frac{\sqrt{k}\omega_3}{\pi\sqrt{1-\nu_i^2 - \omega_2^2}}
[\cos^2\phi_1 I_1 - vd_3 I_2] \ ,
\end{eqnarray}
where
\begin{eqnarray}
I_1 &=& \int \frac{rdr}{\sqrt{r^2 + r_0^2}} = \sqrt{r^2 + r_0^2},
\nonumber \\ I_2 &=& \int \frac{dr}{\sqrt{r^2 + r_0^2}} = \log( r
+ \sqrt{r^2 + r_0^2}) \ .
\end{eqnarray}
Now we can combine different charges to find a relationship among themselves,
firstly combining $E$ and $P_i$ we get,
\begin{equation}
E - \frac{P_i}{\nu_i} = \frac{\sqrt{k}}{\pi\sqrt{1-\nu_i^2 -
\omega_2^2}}v(c_i - c_0)I_2 \ ,
\end{equation}
then combining the $J_{\phi}$'s we get,
\begin{equation}
\frac{J_{\phi_2}}{\omega_2} + \frac{J_{\phi_3}}{\omega_3} =
\frac{\sqrt{k}}{\pi\sqrt{1-\nu_i^2 - \omega_2^2}}[I_1 - v(d_2 +
d_3)I_2] \ ,
\end{equation}
and combining these two relations we get,
\begin{equation}
\sum_{j=2,3}\frac{J_{\phi_j}}{\omega_j} +
\frac{\sum_jd_j}{c_i - c_0} \Big(E - \frac{P_i}{\nu_i}\Big) =
\frac{\sqrt{k}}{\pi\sqrt{1-\nu_i^2}}I_1 \ . \label{orb 2}
\end{equation}
Note that $I_1$ is constant once we fix the integration limit and
it will be finite unless $r \to \infty$. So, this relation
(\ref{orb 2}) differ from the relation of D3-branes (\ref{orb 1})
by a constant factor.
\section{Strings in D1-brane background}
In this section we proceed to the study of the rotating and
orbiting string solutions of the F-string in D1-brane background.
In the case, we have from equation (\ref{eom 2}),
\begin{eqnarray}
&& (1-v^2)^2\frac{\partial}{\partial y}\Big[\frac{1}{r}
\cos^2\phi_1\cos^2\phi_3\frac{\partial \phi_5}{\partial y}\Big] =
\frac{1}{r}\cos^2\phi_1\cos^2\phi_3 \sin\phi_5\cos\phi_5 \nonumber
\\ &\times&  \Big[\frac{\omega_6^2d_6^2r^2}{\cos^4\phi_1
\cos^4\phi_3\sin^4\phi_5} - \frac{\omega_7^2d_7^2
r^2}{\cos^4\phi_1\cos^4\phi_3 \cos^4\phi_5} + \omega_7^2 -
\omega_6^2 \Big] \ , \label{eom 24}
\end{eqnarray}
then from equation (\ref{eom 3}),
\begin{eqnarray}
&& (1-v^2)^2\frac{\partial}{\partial y}\Big[\frac{1}{r}
\cos^2\phi_1\frac{\partial \phi_3}{\partial y}\Big] =
\frac{1}{r}\cos^2\phi_1\sin\phi_3\cos\phi_3
\Big[\frac{\omega_4^2d_4^2r^2}{\cos^4\phi_1 \sin^4\phi_3}
\nonumber \\ &-& \frac{\omega_6^2d_6^2r^2}{\cos^4\phi_1
\cos^4\phi_3\sin^2\phi_5} -
\frac{\omega_7^2d_7^2r^2}{\cos^4\phi_1\cos^4\phi_3 \cos^2\phi_5} -
\omega_4^2 \nonumber \\ &+& \omega_6^2\sin^2\phi_5 +
\omega_7^2\cos^2\phi_5   - (1-v^2)^2 \Big(\frac{\partial
\phi_5}{\partial y}\Big)^2  \Big] \ , \label{eom 25}
\end{eqnarray}
then from equation (\ref{eom 4}),
\begin{eqnarray}
&& (1-v^2)^2\frac{\partial}{\partial y}\Big[\frac{1}{r}
\frac{\partial \phi_1}{\partial y}\Big] =
\frac{1}{r}\sin\phi_1\cos\phi_1 \Big[
\frac{\omega_2^2d_2^2r^2}{\sin^4\phi_1} -
\frac{\omega_4^2d_4^2r^2}{\cos^4\phi_1\sin^2\phi_3} \nonumber \\
&-& \frac{\omega_6^2d_6^2r^2}{\cos^4\phi_1\cos^2\phi_3
\sin^2\phi_5} -
\frac{\omega_7^2d_7^2r^2}{\cos^4\phi_1\cos^2\phi_3\cos^2\phi_5} -
\omega_2^2 \nonumber \\ &+& \omega_4^2\sin^2\phi_3 +
\omega_6^2\cos^2\phi_3\sin^2\phi_5 +
\omega_7^2\cos^2\phi_3\cos^2\phi_5 \nonumber \\ &-& (1-v^2)^2\Big(
\Big(\frac{\partial \phi_3}{\partial y}\Big)^2 +
\cos^2\phi_3\Big(\frac{\partial \phi_5}{\partial y}\Big)^2  \Big)
\Big] \ , \label{eom 26}
\end{eqnarray}
and finally from equation (\ref{eom 5}) we get,
\begin{eqnarray}
&& 2(1-v^2)^2\frac{\partial}{\partial y} \Big[\frac{1}{r^3}
\frac{\partial r}{\partial y}\Big] = [1-\frac{c_0^2}{r^6} -
\nu_i^2\{1 - \frac{c_i^2}{r^6}\}]\partial_r(r^3) \nonumber \\ &+&
(1-v^2)^2\Big(\frac{\partial r}{\partial
y}\Big)^2\partial_r(\frac{1}{r^3}) +
\Big[\frac{\omega_2^2d_2^2r^2}{\sin^2\phi_1} +
\frac{\omega_4^2d_4^2r^2}{\cos^2\phi_1\sin^2\phi_3} \nonumber \\
&+& \frac{\omega_6^2d_6^2r^2}{\cos^2\phi_1\cos^2\phi_3
\sin^2\phi_5} +
\frac{\omega_7^2d_7^2r^2}{\cos^2\phi_1\cos^2\phi_3\cos^2\phi_5}
\nonumber \\ &-& \omega_2^2\sin^2\phi_1 -
\omega_4^2\cos^2\phi_1\sin^2\phi_3 -
\omega_6^2\cos^2\phi_1\cos^2\phi_3\sin^2\phi_5 \nonumber \\ &-&
\omega_7^2\cos^2\phi_1\cos^2\phi_3\cos^2\phi_5 + (1-v^2)^2\Big(
\Big(\frac{\partial \phi_1}{\partial y}\Big)^2  +
\cos^2\phi_1\Big(\frac{\partial \phi_3}{\partial y}\Big)^2
\nonumber \\ &+&  \cos^2\phi_1\cos^2\phi_3\Big(\frac{\partial
\phi_5}{\partial y}\Big)^2 \Big)\Big]\partial_r(\frac{1}{r})\ .
\nonumber \\ \label{eom 27}
\end{eqnarray}
This is a much more complicated system than D5 brane as here all the variables $r$, $\phi_1$, $\phi_3$ and $\phi_5$ are coupled and we don't expect to solve them independently as this is a non integrable system.
In the constraint space,
\begin{equation}
-c_0 + \nu_i^2c_i + \omega_2^2d_2 + \omega_4^2d_4 + \omega_6^2d_6
+ \omega_7^2d_7 =0 \ , \label{cnst 18}
\end{equation}
the Virasoro (\ref{Virasoro 1}) becomes,
\begin{eqnarray}
&& (1 - v^2)^2 \Big[\Big(\frac{\partial r}{\partial y}\Big)^2 +
r^2\Big(\frac{\partial \phi_1}{\partial y}\Big)^2 +
r^2\cos^2\phi_1\Big(\frac{\partial \phi_3}{\partial y}\Big)^2 +
r^2\cos^2\phi_1\cos^2\phi_3\Big(\frac{\partial \phi_5}{\partial
y}\Big)^2 \Big] \nonumber \\ &=& (1-\nu_i^2)r^6 + c_0^2 -
\nu_i^2c_i^2  - r^2\omega_2^2\sin^2\phi_1 -
r^2\omega_4^2\cos^2\phi_1\sin^2\phi_3 \nonumber \\ &-&
r^2\omega_6^2\cos^2\phi_1\cos^2\phi_3\sin^2\phi_5 -
r^2\omega_7^2\cos^2\phi_1\cos^2\phi_3\cos^2\phi_5 -
\frac{\omega_2^2d_2^2r^4}{\sin^2\phi_1} \nonumber \\ &-&
\frac{\omega_4^2d_4^2r^4}{\cos^2\phi_1 \sin^2\phi_3} -
\frac{\omega_6^2d_6^2r^4}{\cos^2\phi_1\cos^2\phi_3 \sin^2\phi_5} -
\frac{\omega_7^2d_7^2r^4}{ \cos^2\phi_1\cos^2\phi_3 \cos^2\phi_5}
\ .  \label{Virasoro 9}
\end{eqnarray}
This Virasoro (\ref{Virasoro 9}) suggests that to have some
solutions we need to put three of the variables among  $r$,
$\phi_1$, $\phi_3$ and $\phi_5$ to be constants. As three of the
variables are being constant here we have to deal with much more
constraints. In the following subsection we will discuss one types
of such solutions.
\subsection{Rotating String Solution: Constant ($r$, $\phi_3$, $\phi_5$)}
As we put $r$, $\phi_3$ and $\phi_5$ as constant, these equations
will impose constraints on the system as mentioned previously. In
summary, all the constraints are given by,
\begin{eqnarray}
\omega_2^2 &=& \omega_4^2 = \omega_6^2 = \omega_7^2,
\>\>\>\>\>\>\>\>\>\>\>\>\>\>\>\>\>\>\>\>\>\>\>\>\>\>\>\>
d_6^2\cos^4\phi_5 = d_7^2\sin^4\phi_5, \nonumber \\ 2\omega_2^2
&=& (1-\nu_i^2)r^4 + \frac{c_0^2 - \nu_i^2c_i^2}{r^2},
\>\>\>\>\>\>\> d_4^2\cos^4\phi_3\sin^4\phi_5 = d_6^2\sin^4\phi_3 \
. \label{cnst 19}
\end{eqnarray}
Under all these constraint (\ref{cnst 19}), we have,
\begin{equation}
\frac{\partial \phi_1}{\partial y} = \frac{\omega_2}{1-v^2}
\frac{\sqrt{(\sin^2\phi_1 - \sin^2\phi_{min})(\sin^2\phi_{max} -
\sin^2\phi_1)}}{\sin\phi_1\cos\phi_1} \ ,
\end{equation}
where $\sin^2\phi_{min} + \sin^2\phi_{max} = 1 + r^2(d_2^2 - \frac{d_4^2}{\sin^4\phi_3})$ and $\sin^2\phi_{min}  \sin^2\phi_{max} =  r^2d_2^2 $. In this case the conserved charges are given by,
\begin{eqnarray}
E &=& \pm \frac{\sqrt{k}(r^3-vc_0)}{2\omega_2} \ , \>\>\> P_i =
\pm \frac{\sqrt{k}\nu_i(r^3-vc_i)}{2\omega_2} \ ,
\end{eqnarray}
we can combine them as
\begin{equation}
E - \frac{P_i}{\nu_i} = \pm \frac{\sqrt{k}v(c_i - c_0)}{2\omega_2}
\ . \label{rel 12}
\end{equation}
Note that here we didn't take the sum over $\frac{P_i}{\nu_i}$, because for D1-brane $i=1$. Also,
\begin{eqnarray}
J_{\phi_2} &=& \pm \frac{\sqrt{k}}{4} [\frac{1}{r}(\sin^2\phi_{min} +
\sin^2\phi_{max}) - 2vd_2], \nonumber \\
J_{\phi_4} &=& \pm \frac{\sqrt{k}\omega_4}{4\omega_2}
[ 2(\frac{1}{r}\sin^2\phi_3 - vd_4) - \frac{1}{r}\sin^2\phi_3(\sin^2\phi_{min} + \sin^2\phi_{max})],
\nonumber \\
J_{\phi_6} &=& \pm \frac{\sqrt{k}\omega_6}{4\omega_2} [ 2(\frac{1}{r}\cos^2\phi_3\sin^2\phi_5 - vd_6) -
\frac{1}{r}\cos^2\phi_3\sin^2\phi_5(\sin^2\phi_{min} + \sin^2\phi_{max})], \nonumber \\
J_{\phi_7} &=& \pm \frac{\sqrt{k}\omega_7}{4\omega_2} [ 2(\frac{1}{r}\cos^2\phi_3\cos^2\phi_5 - vd_7)
- \frac{1}{r}\cos^2\phi_3\cos^2\phi_5(\sin^2\phi_{min} + \sin^2\phi_{max})] \ . \nonumber \\
\end{eqnarray}
We can also combine,
\begin{eqnarray}
&& \frac{J_{\phi_2}}{\omega_2} + \frac{J_{\phi_4}}{\omega_4} +
\frac{J_{\phi_6}}{\omega_6} + \frac{J_{\phi_7}}{\omega_7}   = \pm
\frac{\sqrt{k}}{2\omega_2}[\frac{1}{r} - v(d_2 + d_4 + d_6 + d_7)]
\ . \label{rel 13}
\end{eqnarray}
Combining equation (\ref{rel 12}) and equation (\ref{rel 13}), we get,
\begin{equation}
E - \frac{P_i}{\nu_i} -\sum_{j=2,4,6,7}\frac{J_{\phi_j}}{\omega_j}
=  \pm\frac{\sqrt{k}}{2\omega_2} [v(-c_0 + c_i + \sum_j d_j) - \frac{1}{r}] \ . \label{rot 3}
\end{equation}
As $r$ is constant here, the right hand side of (\ref{rot 3}) is
constant, and the relation is of similar type that we obtained for
D3 and D5 brane background. One will obtain similar kind of
solutions in the cases of (i) $\phi_3$ as variable and $r$,
$\phi_1$ and $\phi_5$ as constants and (ii) $\phi_5$ as variable
and $r$, $\phi_1$ and $\phi_3$ as constants.

\subsection{Orbiting String Solution: Constant $(\phi_1$, $\phi_3$, $\phi_5)$} In this section we study the orbiting strings in the
D1-brane background. In this case $r$ is the only variable and all
angular variable are kept constant. The constraint space is then
given by,
\begin{eqnarray}
&& c_2 = c_0^2 - \nu_i^2c_i^2, \>\>\> \omega_2^2 = \omega_4^2 =
\omega_6^2 = \omega_7^2, \>\>\> d_6^2\cos^4\phi_5 =
d_7^2\sin^4\phi_5,\nonumber \\ && d_4^2\cos^4\phi_3\sin^4\phi_5 =
d_6^2\sin^4\phi_3, \>\>\> d_2^2\cos^4\phi_1\sin^4\phi_3 =
d_4^2\sin^4\phi_1 \ . \label{cnst 22}
\end{eqnarray}
So, in this case we have,
\begin{equation}
\frac{\partial r}{\partial y} = \frac{\sqrt{1 - \nu_i^2}}{1-v^2}
\sqrt{(r^2 - a^2)(r^2 - b^2)(r^2 - c^2)} \ ,
\end{equation}
where $a^2 + b^2 + c^2 = \frac{\omega_2^2d_2^2}{(1 -
\nu_i^2)\sin^4\phi_1}$ and $a^2b^2 + a^2c^2 + b^2c^2 =
-\frac{\omega_2^2}{1 - \nu_i^2}$, and $a^2b^2c^2 = -\frac{c_2}{1 -
\nu_i^2}$. Now the conserved charges in this case are given by,
\begin{eqnarray}
E &=& \frac{\sqrt{k}}{\pi \sqrt{1 - \nu_i^2}} [I_1 -vc_0I_2] ,
\>\> P_i = \frac{\sqrt{k}\nu_i}{\pi \sqrt{1 - \nu_i^2}} [I_1 -vc_i
I_2] , \nonumber \\  J_{\phi_2} &=& \frac{\sqrt{k} \omega_2}{\pi
\sqrt{1 - \nu_i^2}} [\sin^2\phi_1I_3 - vd_2I_2], \>\> J_{\phi_4} =
\frac{\sqrt{k} \omega_4}{\pi \sqrt{1 -
\nu_i^2}} [\cos^2\phi_1\sin^2\phi_3I_3 - vd_4I_2] , \nonumber \\
J_{\phi_6} &=& \frac{\sqrt{k} \omega_6}{\pi \sqrt{1 - \nu_i^2}}
[\cos^2\phi_1\cos^2\phi_3\sin^2\phi_5I_3 - vd_6I_2], \nonumber \\
J_{\phi_7} &=& \frac{\sqrt{k} \omega_7}{\pi \sqrt{1 - \nu_i^2}}
[\cos^2\phi_1\cos^2\phi_3\cos^2\phi_5I_3 - vd_7I_2],
\end{eqnarray}
where
\begin{eqnarray}
I_1 &=& \int \frac{r^3 dr}{\sqrt{(r^2 - a^2)(r^2 - b^2)(r^2 -
c^2)}} \nonumber \\ &=& \frac{\sqrt{(r^2 - b^2)(r^2 -
c^2)}}{\sqrt{r^2 - a^2}} - \sqrt{b^2 - a^2}
E\Big[i\sinh^{-1}\Big(\frac{\sqrt{a^2 - b^2}}{\sqrt{r^2 -
a^2}}\Big), \frac{a^2 - c^2}{a^2 - b^2}\Big] \nonumber \\ &-&
\frac{b^2}{\sqrt{b^2 - a^2}} F\Big[i\sinh^{-1}\Big(\frac{\sqrt{a^2
- b^2}}{\sqrt{r^2 - a^2}}\Big), \frac{a^2 - c^2}{a^2 - b^2}\Big],
\nonumber \\ I_2 &=&  \int \frac{dr}{\sqrt{(r^2 - a^2)(r^2 -
b^2)(r^2 - c^2)}} \nonumber \\ &=& \frac{1}{b\sqrt{c^2 - a^2}}
F\Big[\sin^{-1}\Big(\sqrt{\frac{r^2(c^2 - a^2)}{c^2(r^2 -
a^2)}}\Big), \frac{(a^2 - b^2)c^2}{b^2(a^2 - c^2)}\Big], \nonumber
\\ I_3 &=&  \int \frac{dr}{\sqrt{r(r^2 - a^2)(r^2 - b^2)(r^2 -
c^2)}} \nonumber \\ &=& \frac{1}{c^2\sqrt{a^2 - c^2}}
\Pi\Big[1-\frac{b^2}{c^2}, \sin^{-1}\Big(\sqrt{\frac{r^2 -
c^2}{b^2 - c^2}}\Big), \frac{b^2 - c^2}{a^2 - c^2}\Big]  \ ,
\end{eqnarray}
where $F(\varphi,m)$, $E(\varphi,m)$ and $\Pi(n,\varphi,m)$ are
the incomplete elliptic integrals of first, second and third kind
respectively. We have to find out which integration limits are
useful in this case. However, we can combine different charges to
find a relationship among themselves, firstly combining $E$ and
$P_i$ we get,
\begin{equation}
E - \frac{P_i}{\nu_i} = \frac{\sqrt{k}}{\pi\sqrt{1-\nu_i^2}}v(c_i
- c_0)I_2 \ ,
\end{equation}
then combining the $J_{\phi}$'s we get,
\begin{equation}
\frac{J_{\phi_2}}{\omega_2} + \frac{J_{\phi_4}}{\omega_4} +
\frac{J_{\phi_6}}{\omega_6} + \frac{J_{\phi_7}}{\omega_7} =
\frac{\sqrt{k}}{\pi\sqrt{1-\nu_i^2}}[I_3 - v(d_2 + d_4 + d_6 +
d_7)I_2] \ ,
\end{equation}
and combining these two relations we get,
\begin{equation}
    \sum_{j=2,4,6,7}\frac{J_{\phi_j}}{\omega_j} +
\frac{\sum_j d_j}{c_i - c_0} \Big(E -
\frac{P_i}{\nu_i}\Big) =\frac{\sqrt{k}}{\pi\sqrt{1-\nu_i^2}}I_3 \ . \label{orb 3}
\end{equation}
Unless we find the value of $I_3$ using appropriate integration
limit, we can not comment on the nature of this relation (\ref{orb
3}) in detail. However, this relation seems to be different from
(\ref{orb 1}) and (\ref{orb 2}) of D3 and D5-brane backgrounds.
\section{Conclusion}
This paper is devoted to study various rigidly rotating and
orbiting solutions of the F-string in various Dp-brane backgrounds
in type IIB string theory. As the string equations of motion in
general Dp-brane backgrounds are non integrable (except for D3
branes), we do not except to find solutions in full generality. We
investigate various subspaces of Dp-brane background and found two
kinds of solutions, namely the rotating string and the orbiting
string solutions. For the D3-brane background we find two classes
of rotating string solutions. First one is the well known giant
magnon and single spike solution. Second one is a new kind of
rotating string solution for which $E - J = $ constant for constant
$E$ and $J$. In fact it is the second kind of solution which has
 similar structure for all the D-brane
backgrounds we have considered here. Further we have found another
class of solutions which are the so called orbiting string
solutions. These are the string configurations where the string
orbit around the core of Dp-brane. It would be interesting to
study how orbiting strings differ for different Dp-branes by
putting appropriate integration limits. Looking for the dual gauge
theory operators for these kind of string states will be of
interest. Further one would like to study the rotating string
solution in some intersecting brane backgrounds with mixed flux 
and see if one gets the giant magnon and single spike like solutions. 
It will also be interesting to study the D-string solutions in Dp-brane
background.

\end{document}